\documentstyle[sprocl]{article}

\input{psfig}

\def\be{\begin{equation}}
\def\ee{\end{equation}}
\def\bea{\begin{eqnarray}}
\def\eea{\end{eqnarray}}

\newcommand\x{\mbox{\boldmath $x$}}
\newcommand\ihat{\hat{\imath}}

\begin{document}
\begin{flushright}
HIP-1998-47/TH
\end{flushright}
\vspace{.2cm}

\title{VORTICES IN EQUILIBRIUM SCALAR ELECTRODYNAMICS}

\author{A. RAJANTIE}
\address{Helsinki Institute of Physics, P.O.Box 9, 
00014 University of Helsinki, Finland}

\author{K. KAJANTIE, M. LAINE}
\address{Theory Division, CERN, CH-1211 Geneva 23, Switzerland}

\author{M. KARJALAINEN}
\address{Department of Physics, P.O.Box 9, 
00014 University of Helsinki, Finland}

\author{J. PEISA}
\address{Department of Physics, University of Wales Swansea,
Singleton Park, Swansea~SA2~8PP, United Kingdom}


\maketitle\abstracts{ 
Scalar electrodynamics can be used to investigate the formation of
cosmic strings in the early universe. We present the results of lattice 
Monte Carlo simulations of an effective three-dimensional U(1)+Higgs 
theory that describes the equilibrium properties of finite-temperature
scalar electrodynamics 
near the transition. A gauge-invariant criterion for the existence of 
a vortex is used in measuring the properties of the vortex 
network in the equilibrium state both in the Coulomb and in the 
Higgs phase of the system. The naive definition of the vortex density
becomes meaningless in the continuum limit and special care is needed
in extracting physical quantities. Numerical evidence for a
physical discontinuity in the vortex density is given.}

The traditional picture for defect formation in gauge theories relies
on the Kibble mechanism~\cite{ref:KibbleMech}: 
the effective potential
changes shape at the critical point and the field relaxes to different
minima in space-time points separated by more than a
correlation length. However, 
strictly speaking
this approach can only be applied to transitions in which a symmetry
breaks spontaneously, and gauge symmetries cannot be 
broken~\cite{ref:Elitzur}. In some cases, e.g.~in the electroweak theory,
the transition line ends at a critical point after which there
is only an analytical crossover between the 
phases~\cite{ref:Crossover}. It is clear that
some changes must be made to the picture of defect formation in
cases like this.

We will here present the results of the simulations of
three-dimensional scalar electrodynamics,
i.e.~the Abelian Higgs model.
The details can be found in the original 
publications~\cite{ref:lammi,ref:plb}. 
The theory describes the equilibrium properties
of finite-temperature 
relativistic scalar electrodynamics near the transition~\cite{ref:dimred}, 
and has the same form as the
Ginzburg-Landau theory of superconductivity. 
We use lattice Monte Carlo simulations to 
study the behavior of thermally generated vortices, 
i.e.~line-like topological defects, in the
two phases of the theory. Similar studies have previously been 
carried out for
the globally symmetric theory using the Langevin equation
\cite{ref:antunes}.
Our final aim is to understand non-perturbatively the process of
defect formation, but we find it necessary to understand the
equilibrium behavior first.

The continuum theory is given by the action
\be
S=\int d^3x\biggl[\frac{1}{4} F_{ij}^2+
|D_i\phi|^2 
+y \phi^*\phi + x \left(\phi^*\phi\right)^2\bigg], 
\label{equ:action}
\ee
where $F_{ij}=\partial_iA_j-\partial_jA_i$, $D_i=\partial_i+iA_i$
and $y$ is renormalized in the $\overline{\rm MS}$ scheme at $\mu=1$. 
In this notation all quantities are dimensionless.
If we write $\phi(\x)=v(\x)\exp[i\gamma(\x)]$,
the action is invariant under the gauge transformation
\be
\gamma(\x)\to [\gamma(\x)+\theta(\x)]_\pi,\quad
A_i(\x)\to A_i(\x)-\partial_i\theta(\x),
\label{equ:gauge1}
\ee
where $[X]_\pi\equiv X+2\pi n$ such that $[X]_\pi\in (-\pi,\pi]$.
The relations of $x$ and $y$ to the parameters
of the 4D scalar electrodynamics and superconductors, as well as the
determination of the phase diagram, have been discussed 
elsewhere~\cite{ref:u1big}.

\begin{figure}

\psfig{file=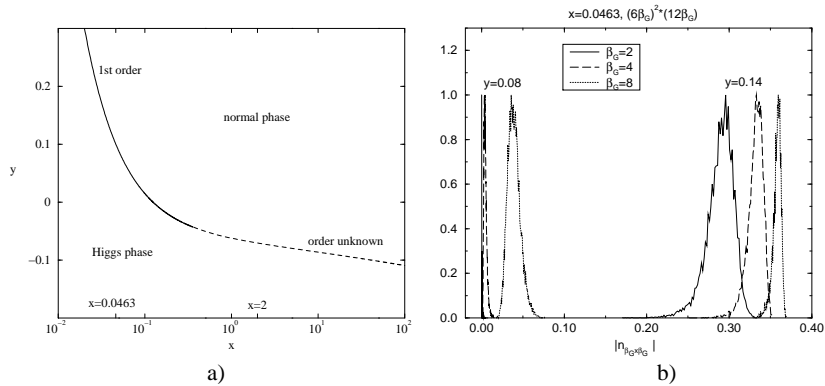,width=4.6in}

\vspace{-.2cm}
\caption{
\label{fig:kuva}
a) The phase diagram of the system.
b) The histogram of the volume averages of the winding number of a 
curve of constant size in physical units at various values of
the lattice spacing $a=1/\beta_G$. In the continuum limit
$\beta_G\to\infty$ the distance of the peaks seems to approach
a finite value, which is expected to be independent of the
regulator.}

\end{figure}

The phase diagram of the system consists of the Coulomb and the Higgs
phase.
Although a mean-field analysis suggests a symmetry-breaking 
transition, it turns out that there is, in fact, no
local order parameter. The mass of the photon acts as a non-local
order parameter, being non-zero in the Higgs phase and
vanishing in the Coulomb
phase~\cite{ref:BorgsNill}. At small $x$, the transition is of first order
as predicted by perturbation theory,
but at some critical value of $x$ it becomes continuous~\cite{ref:u1big}
(See Fig.~\ref{fig:kuva}a).


For numerical simulations, the theory must be defined on a lattice.
We use the non-compact formulation, which means that there is a
real number $\alpha_i(\x)$ corresponding to the continuum gauge field 
$A_i(\x)$ on each link $(\x,\x+\ihat)$ between the lattice sites.
On each site there is a scalar field $\phi(\x)=v(\x)\exp[i\gamma(\x)]$.
The lattice analogue of the gauge transformation (\ref{equ:gauge1}) is
\be
\gamma(\x)  \to  [\gamma(\x)+\theta(\x)]_\pi,\quad
\alpha_i(\x) \to 
\alpha_i(\x)+\theta(\x)-\theta(\x+\ihat). \label{equ:gauge2}
\ee
The parameters appearing in the lattice action differ from the 
continuum ones, but the relation can be calculated exactly with a
2-loop computation in lattice perturbation theory~\cite{ref:contlatt}.

To find the vortices, we define for each link the 
quantity~\cite{ref:lammi,ref:plb,ref:ranft}
\be
\label{equ:link}
Y_{(x,x+\hat\imath)}=[\alpha_i(\x)+\gamma(\x+\hat\imath)-
\gamma(\x)]_\pi-\alpha_i(\x). 
\ee
Taking a sum around a closed curve $C$
gives the winding number $n_C$:
\be
\label{equ:winding}
Y_C=\sum_{l\in C}Y_l\equiv 2\pi n_C.
\ee
The winding number is a gauge-invariant integer and gives the number of
vortices going through the curve 
$C$~\footnote{The standard method of locating vortices by finding the
zeros of the Higgs field \cite{ref:popov} has often been used
also in gauge theories \cite{ref:yates}. However, on a lattice
one can get rid of all the zeros by choosing e.g.~the unitary gauge.
Even if some other gauge is used, the physical interpretation of the vortices
found this way is ambiguous.}.
Using the difference of only the phase angles in
Eq.~(\ref{equ:link}) 
would lead to a non-invariant quantity.

At the mean-field level the notion of a vortex makes only sense in the
Higgs phase, but our definition (\ref{equ:winding}) is perfectly
valid in all phases and agrees with the intuitive picture of a vortex
whenever the bare tree-level potential has a degenerate minimum.
We calculate the full path integral numerically using lattice Monte
Carlo simulations, which means including the effect of thermal fluctuations
to the mean-field picture.

The quantity we are mainly interested in at 
this stage is the density of thermally generated vortices. The naive way 
to calculate it is to
take the absolute value of the winding number of a single plaquette
and measure its expectation value. However, it turns out that
in the continuum limit, this quantity approaches a universal quantity
$\approx 0.2$, which is independent of the parameters of
the theory~\cite{ref:plb}. 
The reason is that the winding number of a single plaquette
is an ultraviolet quantity, and the ultraviolet behavior of the
theory is given by a free massless complex scalar field.

To obtain a physical quantity, we take a square curve $C$ and keep its
size constant in physical units as we approach the continuum limit.
While this quantity contains a lot of ultraviolet noise, there is
also a physical contribution, and the problem is to extract it. 
An analogous quantity is $\langle\phi^*\phi\rangle$: it diverges in
the continuum limit, but the divergent term is constant and can be 
calculated exactly~\cite{ref:contlatt}. 
The remaining finite part can then be used to probe the phase diagram 
of the theory. For $\langle |n_C| \rangle$ we cannot subtract the
divergence exactly, but if it is analytical in the parameters $x$ and $y$
as we expect,
the difference in $\langle |n_C| \rangle$ above and below the
transition line is a physical quantity~\cite{ref:plb}. Some numerical 
evidence for that is shown
in Fig.~\ref{fig:kuva}b.

In practice, it is rather difficult to extract the physical contents of the
vortex density defined here.
However, it shows that in the continuum limit, more care is needed
than simply interpreting the plaquettes with non-zero winding numbers
as physical vortices.
One has to come
up with new observables that give better understanding of
the problem before a non-perturbative
understanding of defect formation in gauge theories can be obtained.

\end{document}